\documentclass[graybox]{svmult}

\usepackage{type1cm}        
\usepackage{makeidx}         
\usepackage{graphicx}        
\usepackage{multicol}        
\usepackage[bottom]{footmisc}

\usepackage{newtxtext}      
\usepackage[varvw]{newtxmath}      
\usepackage{cases}
\usepackage{amsmath}
\usepackage{amsfonts}

\def\pa{\partial\Omega}

\def\R{{\mathbb R}}

\newcommand{\be}{\begin{equation}}
\newcommand{\ee}{\end{equation}}

\makeindex          

\begin{document}

\title*{Search and escape of mortal random walkers}
\author{E. Abad and S. B. Yuste}
\institute{E. Abad \at Universidad de Extremadura; Departamento de Física Aplicada and Instituto de Computación Científica Avanzada (ICCAEx), Mérida, Spain, \email{eabad@unex.es}
\and S.B. Yuste \at Universidad de Extremadura, Departamento de Física and Instituto de Computación Científica Avanzada (ICCAEx), Badajoz, Spain, \email{santos@unex.es}}

\maketitle

\abstract{We review some representative results for first-passage problems involving so-called mortal or evanescent walkers, i.e., walkers with a finite lifetime. The mortality constraint plays a key role in the modeling of many real scenarios, as it filters out long Brownian trajectories, thereby drastically modifying space exploration properties. Among such scenarios, we consider here different first-passage problems, including one or many searchers, resetting, anomalous diffusion, evolving domains, and the narrow escape problem. In spite of the different physics, the mathematical treatment draws strongly on the formalism for standard (i.e., immortal) walkers.}

\section{Introduction}

Mortal or evanescent walkers ``die'' at a certain rate as they move. In the context of search and escape problems, the death process should be understood as any constraint leading to a premature end of a diffusive trajectory, or as a change of identity which prevents the walker to find a target or to exit a domain. Beyond the not so frequent cases of literal death (e.g., a prey hunted by a predator), other constraints that can be regarded as death processes in a broad sense include degradation, inactivation, premature detachment from a substrate, decay into an inert species, or absorption/trapping of any kind.  In spite of the ubiquity of such mechanisms and of their drastic effects, they are not always incorporated in random walk models, or not always in the correct way. For example, in models of anomalous diffusion, mortality typically leads to a transport term that includes the mortality rate (see e.g. Eq.~\eqref{subdiffusion-decay} below); this peculiarity has been sometimes overlooked, as it is at first glance antiintuitive. 

An example of mortal walks often invoked in the literature concerns the diffusion of radioactive particles, which decay according to Rutherford-Soddy’s exponential law~\cite{Nicolaysen57,  Silverman17}.
Other phenomena which can be modeled as diffusion-decay processes include motility of spermatozoa~\cite{Meerson15}, surface diffusion on catalytic substrates~\cite{Abad15b}, scavenging reactions~\cite{Sano79}, photoluminescence decay due to electron-hole recombination~\cite{Seki05}, diffusion of defects~\cite{Yuste2013}, mRNA translation~\cite{Grebenkov17}, light absorption in tissues~\cite{Bonner87}, morphogen gradient formation~\cite{Wartlick09, Yuste10}, transport and detachment of molecular motors from polymer tracks~\cite{Kolomeisky00}, virus trafficking subject to killing~\cite{Holcman2015}, intracellular signaling~\cite{Ma20}, and channel transport of degrading information molecules~\cite{Briantceva23}, to name but a few. Even the random walk corresponding to the classical problem of the gambler’s ruin can be modified by allowing for a finite probability to quit the game before each bet~\cite{Abad15b}, resulting in a strong shortening of the game duration.

A fast mortality results in a drastic decrease of long trajectories; consequently, first-passage properties conditioned on survival may be very different from their counterparts for immortal walkers. As pointed out by Lawley~\cite{Lawley21}, ``if inactivation is fast, then the conditional first-passage time (FPT) compared to the FPT without inactivation is (i) much faster, (ii) much less affected by spatial heterogeneity, and (iii) much less variable''. Thus, mortality selects the most direct trajectories to a target and shortens the measured search time at the expense of having a poorer statistics; in this sense, it is different from other mechanisms such as e.g. chemotaxis or the Adam-Delbr\"uck reduction-of-dimensionality scenario~\cite{Bala19, Grebenkov22}.

On a more technical note, we also mention that death processes can also be used to ``heal'' some pathologies of certain equations, e.g. the lack of robustness of subdiffusive fractional equations as far as their stationary distribution is concerned~\cite{Fedotov13}. There are also some surprising connections between the mortal walk formalism and averaging dynamics on graphs~\cite{Sikder21}, as well as between lattice Green functions of immortal walkers~\cite{Hughes95} and space exploration properties of mortal ones~\cite{Hughes95, Yuste2013, Yuste14}.

Our aim is to give a short overview of a few representative problems of interest. For reasons of space, it is unfortunately not possible to describe (even minimally) all the problems with mortal walkers studied so far by different authors (and by ourselves). Moreover, the description of those presented here necessarily remains schematic. A previous book chapter~\cite{Yuste14} coauthored by two of us also deals with mortal random walkers, although mainly on discrete supports and for a different set of problems. For the sake of continuity with Ref.~\cite{Yuste14}, we begin our presentation with the discrete case, but then focus exclusively on continuum diffusion.

\section{Basic properties of mortal Brownian motion}

\subsection{Probability to be at a certain location }

We consider a mortal  walker performing a discrete-time Markovian random walk consisting of equiprobable jumps to nearest-neighbor sites on a connected graph. This walker has a probability $q$ of surviving at each time step; thus, its probability of demise at each time step is $1-q$. Standard (``immortal'') random walks are recovered for $q = 1$.

We will devote the remainder to the continuum or diffusion limit of a mortal random walk, as this has been extensively studied in the literature. The lack of explicit spatial structure (apart from dimensionality and boundary conditions) makes this limit generally easier to analyze. We note, however, that discrete geometries are important to mimic the details of real systems, and a number of works address this case by making use of lattice Green functions~\cite{Lohmar2009, Yuste2013, Abad13b, Abad15b, Re15, Esguerra17, Abad18}, as well as of recursive relations in the case of hierarchical lattices~\cite{Bala19} and fractal supports~\cite{Bala19, Wu20}.

For concreteness, we focus on the case of an infinite one-dimensional lattice and label its sites with integers. We respectively denote by $a$ and $\tau$ the lattice constant and time step. The probability $p_{j}(n)$ that the walker is at site $j$ at time $n\tau$ reads
\be
p_{j}(n+1) = \tfrac{1}{2} q\,[p_{j-1}(n) + p_{j+1}(n)].
\label{1drw}
\ee
We build the continuum limit of this difference equation in the usual way, i.e., by  letting $a\rightarrow 0$ and  $\tau \rightarrow 0$.
Additionally we let $q \rightarrow 1$ in such a way that~\cite{Bala19}
\be
\lim \, q a^{2}/(2\tau) = D,\qquad \lim\, (1-q)/\tau = \mu,
\label{difflimit}
\ee
where $D$ stands for the walker's diffusivity and $\mu$ for the mortality rate. Replacing $ja$ by $x$ and $n\tau$ by $t$ in Eq.~\eqref{1drw}, and retaining the symbol $p(x,t)$ for the positional pdf, we obtain the following diffusion-decay equation:
\be
\frac{\partial p}{\partial t} = D\,\frac{\partial^{2}p}{\partial x^{2}} - \mu \,p.
\label{diffusioneqn}
\ee
The solution $p(x,t \vert\, x_0)$ of Eq.~\eqref{diffusioneqn} satisfying the initial condition $p(x,0) = \delta (x-x_{0})$
reads
\be
p(x,t \vert\,x_0) = e^{-\mu t}\, p_I(x,t\vert\,x_0),
\label{diffeqnfundsoln}
\ee
where 
\be
p_I(x,t \vert\,x_0) =   (4\pi D t)^{-1/2}\,e^{-(x-x_{0})^{2}/(4Dt)}
\label{pIxt}
\ee
denotes the propagator for immortal Brownian motion. This is just the standard Gaussian solution with the extra exponentially decaying factor $e^{-\mu t}$. A similar decomposition holds true in $d$ dimensions, so the corresponding $d$-dimensional solution $p(\mathbf{x},t\vert\,\mathbf{x}_0)$ is given by~\eqref{diffeqnfundsoln} and~\eqref{pIxt}  replacing $(4\pi D t)^{-1/2}$ by $(4\pi D t)^{-d/2}$  and $x-x_{0}$ by $|\mathbf{x}-\mathbf{x}_0|$. Note that now $p(\mathbf{x},t \vert\, \mathbf{x}_0)$ is no longer normalized to unity, as $\int \!d\mathbf{x}\,p(\mathbf{x},t\vert\,\mathbf{x}_0)= e^{-\mu t}$. 

\subsection{Probability to die at a certain location}

The probability distribution $p_\mu(x,t \vert\, x_0)$ for a mortal Brownian particle starting at position $x_0$ on the real line to die inside an interval $(x,x+dx)$ between the times $t$ and $t+dt$ is easily obtained by exploiting the statistical independence of the Brownian motion and the death process. Thus, $p_\mu(x,t \vert\, x_0)=p_\mu(t) p_I(x,t \vert\, x_0)$, where $p_\mu(t)=\mu e^{-\mu t}$ is the death time distribution. The probability to die at any time at a given location is then obtained by integration~\cite{Meerson19}: 
\be
p_\mu(x \vert\, x_0)=\int_0^\infty p_\mu(x,t \vert\, x_0)\,dt=\sqrt{\frac{\mu}{4D}}e^{-\sqrt{\mu/D}
|x-x_0|}
\ee
This result can be generalized to higher Euclidean dimensions. We quote the two- and three-dimensional results~\cite{Meerson19}:
\begin{numcases}
{p_\mu(\mathbf{x} \vert\, \mathbf{x}_0)=}
\label{2d}
\frac{\mu}{2\pi D}K_0\left(\sqrt{\frac{\mu}{D}}|\mathbf{x}-\mathbf{x}_0|\right), & $d=2$, \\
\label{3d}
\frac{\mu}{4\pi D |\mathbf{x}-\mathbf{x}_0|}\,e^{-\sqrt{\frac{\mu}{D}}|
\mathbf{x}-\mathbf{x}_0|}, 
& $d=3$, 
\end{numcases}
where $K_0$ is the zero-th order modified Bessel function of the second kind. These results differ strongly from the Gaussian profile of the propagator for an immortal walker. As noted by Meerson~\cite{Meerson19}, the expressions given by Eqs.~\eqref{2d} and~\eqref{3d} both diverge at $\mathbf{x}_0$ (but are of course integrable), reflecting the strong localization about the starting point induced by mortality.

\section{First-passage properties of a single searcher}
\label{singlesearcher}

\subsection{Survival probability, conditional first-passage time, and territory explored}
\label{FPTsingle}

Consider a mortal walker (searcher) diffusing from the starting point $x_{0}$ to an 
arbitrary target point $x$ on the real line (we take $x< x_{0}$ without loss of generality). In the remainder, we will occasionally refer to this setting as the one-sided target problem (OSTP). The corresponding FPT density $f(t,x\vert x_{0})$ follows from the Markovian renewal equation
\be
p(x_{1}, t \,\vert\,   x_{0}) = \int_{0}^{t}\!dt^{\prime}\,f(t^{\prime}, x \,\vert \,x_{0})\,
p(x_{1}, t-t^{\prime}\,\vert\, x),
\label{renewaleqn}
\ee
where $x_{1} < x < x_{0}$. With the help of Laplace transforms, we find
\be
f(t,x\,\vert \,x_{0}) =
\frac{e^{-\mu t}}{(4\pi D t^{3})^{1/2}}\,(x_{0}-x)\,e^{-(x_{0}-x)^{2}/(4Dt)}.
\label{continuumfptd}
\ee
The attenuation factor $e^{-\mu t}$ forces the first moment of $f(t,x\,\vert \,x_{0})$ to become finite, whence the non-recurrence of this one-dimensional mortal walk follows. Consequently, first-passage properties are strongly modified, notably the FPT density and the volume explored by the walker (see below). For completeness, we note that Eq.~\eqref{continuumfptd} also follows from the method of images upon use of Eq.~\eqref{diffeqnfundsoln} (see, e.g.,  Eqs. (1) and (2) of Ref.~\cite{Meerson15}).  The probability $F(t, x\,\vert \,x_{0})=\int_0^t dt' f(t',x\,\vert \,x_{0})$ that the searcher has reached point $x$ by time $t$ is~\cite{Abad13, Meerson15}
\begin{align}
\label{exit-t}
 F(t, x\,\vert \,x_{0})&= \tfrac{1}{2} \,
 e^{\sqrt{\frac{\mu}{D}}\,(x_0-x)} \mathrm{erfc}\big([(x_0-x)/\sqrt{4Dt}]+\sqrt{\mu
  t}\big) \nonumber \\
  & +\tfrac{1}{2} \,e^{-\sqrt{\frac{\mu}{D}}\,(x_0-x)}\,
  \mathrm{erfc}\big([(x_0-x)/\sqrt{4Dt}]-\sqrt{\mu t}\big)\,
\end{align}
with $ \mathrm{erfc}(z)=1-\mathrm{erf}(z)$, where $\mathrm{erf}(z) =(2/\!\sqrt{\pi}) \int_0^z e^{-u^2} du$ is the error function.  
The probability that a searcher starting from $x_{0}$ ever reaches $x$ is
\be
{\cal F}(x\,\vert \,x_{0})=\int_{0}^{\infty}\!dt\,f(t,x\,\vert \,x_{0}) =
 \exp\,[- (\mu/D)^{1/2}\,(x_{0}-x)],
\label{continuumtrapprob}
\ee
which exhibits exponential attenuation with increasing distance $x_{0}-x$.  Correspondingly, the (conditional) mean first-passage time (MFPT) reads
\be
\langle T(x\,|\,x_{0}) \rangle =
\frac{\int_{0}^{\infty}\!dt\,t\,f(t,x\,\vert \,x_{0})}
{\int_{0}^{\infty}\!dt\,f(t,x\,\vert \,x_{0})}
= \frac{x_{0}-x}{\sqrt{4 D \mu}}\,.
\label{continuummfpt}
\ee
As in ballistic motion, the mean travel time is thus proportional to the covered distance. Here, the effective speed is $(4D\mu)^{1/2}$. However, one should bear in mind that the conditional MFPT is an average over a finite fraction of trajectories only, namely, those involving a first passage from $x_{0}$ to $x$. This fraction decreases exponentially with increasing distance $x_{0}-x$, implying a loss of statistical accuracy in a hypothetical experimental measurement of the MFPT.

The calculation of the variance of the FPT is also straightforward. We skip the details and give directly the final result (see e.g. Eq. (5b) of Ref.~\cite{Meerson15}):
\be
\sigma^2_T\equiv\langle T^2(x\,|\,x_{0}) \rangle -\langle T(x\,|\,x_{0}) \rangle^2= 
\frac{2D(x_0-x)}{(4D\mu)^{3/2}}.
\ee
The non-recurrence of mortal Brownian motion reduces strongly the explored territory; the latter would be infinite for immortal particles, but turns out to be finite for mortal ones. Formally, the explored volume in the continuum is known as ``the Wiener sausage''.  The characterization of its statistical properties is key for the understanding of diffusion-controlled kinetics and of the so-called target problem~\cite{Yuste2013} in particular. The average volume $v_t$ of the Wiener sausage generated by a mortal Brownian particle of radius $R$ up to time $t$ has been recently determined~\cite{Yuste2013} under the assumption that $t=n\ell^2/(2dD)$ is long enough ($n\gg 1$), thus ensuring that the volume explored by the particle is much larger than its own volume. As it turns out,  $v_t \sim \gamma_d (\ell/R)^2 R^d S_n$ for $d\ge 2$; here, $\gamma_d$ denotes a constant that depends on dimension (e.g., $\gamma_2=1$, $\gamma_3\simeq 2.917$ ~\cite{Berezhkovskii1989}), and $S_n$ is the territory covered by a particle that evolves in a $d$-dimensional simple cubic lattice with period $\ell \ll R$. Explicit expressions for $S_n$ and its final value $S_\infty$ (the latter being of special relevance for applications) can be found in Ref.~\cite{Yuste2013}. E.g., for $d=1$ one finds 
\be
S_\infty=\left[\frac{1+e^{-\tilde \mu}}{1-e^{-\tilde \mu}}\right]^{1/2}, 
\ee
where $\tilde \mu=[\ell^2/(2D)]\mu$. For $d\ge 2$ and small $\mu$, $S_\infty$ goes as $1/\mu$, albeit with a logarithmic correction in the two-dimensional case~\cite{Yuste2013}.

\subsection{First passage with resetting}

Resetting changes drastically first-passage properties~\cite{Evans2011} and has a large number of applications~\cite{Evans20}.
Following~\cite{Belan18}, let us discuss the OSTP (defined in Sect.~\ref{FPTsingle}) 
with a mortal searcher starting from $x_{0} \geq 0$. The searcher is subject to stochastic reset to the initial position $x_{0}$ at a rate $r$. The search process ends either with detection of a point target at $x=0$, or else with the searcher's death. The searcher's pdf $p(x, t \vert\, x_0)$ satisfies the equation~\cite[Supplementary material]{Belan18}
\be
\frac{\partial p}{\partial t}=D \frac{\partial^2 p}{\partial x^2}-(\mu+r) p+r \,\delta\left(x-x_{0}\right) \int_{0}^{\infty} dy \, p(y, t), 
\ee
which must be solved with the initial condition $p(x, 0\vert\, x_0)=\delta\left(x-x_{0}\right)$ and the boundary condition $p(0, t\vert\, x_0)=0$. The second term on the right-hand side takes into account the negative probability flux $(\mu+r)p$ out of each point $x$ due to both resetting and particle death. The third term gives the probability influx at $x_0$ due to resetting of the particles that have survived by time $t$.
For the Laplace transform $\tilde{p}(x,s\vert\, x_0)=\int_{0}^{\infty} dt \, e^{-s t} p(x, t\vert\, x_0) $, we obtain
\be
s \tilde{p}-\delta\left(x-x_{0}\right)=D \frac{\partial^2\tilde{p}}{\partial x^2}-(\mu+r) \tilde{p}+r \, \delta\left(x-x_{0}\right) \int_{0}^{\infty} dy \, \tilde{p}(y,s).
\ee
The pdf $p(x,t \vert\, x_0)$ must fulfil the zero boundary condition at $x=0$, but also remain finite as $x \rightarrow+\infty$. Together with the continuity condition at $x_{0}$, this yields~\cite{Belan18}
\be
\tilde{p}(x,s\vert\, x_0)= \begin{cases}A \sinh \gamma_{s} x, & x \leq x_{0}, \\ A e^{\gamma_{s}\left(x_{0}-x\right)} \sinh \gamma_{s} x_{0}, & x\geq x_{0},\end{cases}
\ee
where $\gamma_{s}=\sqrt{(s+r+\mu) / D}$. The unknown coefficient $A$ is obtained from the jump of the derivative $\partial_{x} \tilde{p}$ at $x=x_{0}$, which gives
$A=\gamma_{s}/[(s+\mu) \exp(\gamma_{s} x_{0})+r]$.

We are now in the position to compute the Laplace transform of the flux $j(t\vert\, x_0)=D \partial_{x} p(x, t\vert\, x_0)$ to the target, which reads 
\be
\tilde{j}(s\vert\, x_0)=D\left.\frac{\partial \tilde{p}}{\partial x}\right|_{x=0}=\frac{s+r+\mu}{(s+\mu) e^{\gamma_{s} x_{0}}+r}.
\ee
From here, the probability to ever reach the target follows as 
\be
p_{r}(x_0)=\int_{0}^{\infty} j(t\vert\, x_0) d t=\tilde{j}(0\vert\, x_0)=\frac{r+\mu}{\mu e^w+r},
\label{detectprob}
\ee
where   $w=\sqrt{(r+\mu) / D} x_{0}$. One then concludes from Eq.~\eqref{detectprob}  that $p_{r}$ decreases monotonically with increasing $r$ if $\mu \geq \mu_{0}=\left(z^{*}\right)^{2} D / x_{0}^{2}$, where $z^{*} \approx 1.59362426 \ldots$ is the solution to $z / 2=1-e^{-z}$.  In contrast, for $\mu <\mu_{0}$, the probability $p_{r}$ takes its maximum at the non-vanishing restart rate $r_{0}=\mu_{0}-\mu$. 

As explained in Ref.~\cite{Belan18}, the observed behavior of $p_{r}$ can be qualitatively understood as follows. When $\mu$ is small compared to $D /x_{0}^{2}$, the searcher lives long enough to perform a typical diffusive path and reach the target. On the other hand, the searcher can still make distant excursions into empty regions of the search space. In this context, a nonvanishing restart rate (resetting) censors fatal paths and increases the chance of finding the target. However, an excessively large resetting rate may hinder target detection, as the searcher has less time between restarts to reach the origin while facing the same mortality rate. Hence, there exists an optimal non-zero resetting rate $r^{*}$ which maximizes the probability of finding the target before death. At this point, it is worth noting that  a recent work by Radice~\cite{Radice23} extends the analysis to other quantities, including explicit expressions for the mean and the variance of the search time.

According to Belan~\cite{Belan18}, the above problem of target detection by a mortal walker subject to resetting can be viewed as a specific type of  Bernoulli experiment with restart (target detection is identified with ``success'', whereas non-detection due to the searcher's premature death is identified with ``failure''). However, Belan's discussion goes beyond this particular example; he concludes that in a generic first-passage process for which an optimal resetting rate $r^{*}$ exists  [in the sense that $r^*$ maximizes (minimizes) the probability of success (failure)],  then for $r=r^{*}$  the completion time of the process conditioned on a specific outcome (no matter whether success or failure) equals the completion time over both successful and failed trials (unconditional MFPT).

\section{$N$-walker problem: order statistics} 

\subsection{Shortest first-passage time}
\label{ShortestFPT}

Let us consider a collection of $N$ (non-interacting) diffusing particles (also termed indistinctly ``walkers'' or ``searchers'' hereafter) that compete to reach a target. In this context, the FPT statistics of the fastest searcher (first searcher to reach the target) is of particular interest. 
Beyond biological problems like gene regulation~\cite{Harbison04}, calcium signaling~\cite{Schuss19} or oocyte fertilization~\cite{Meerson15}, the order statistics of random walkers also plays a key role in other fields, e.g. for information spreading in networks~\cite{Wang08}.

We consider the simplest initial condition, i.e., $N$ mortal searchers all starting at the same distance $x_0>0$ from a target placed at $x=0$. We ask about the FPT of the first one to reach the target. To reduce the problem to a single-particle one (cf. Sect.~\ref{singlesearcher}), we use the simplified notation $f_1(t)\equiv f(t,0\,\vert \,x_0)$ and $F_1(t)\equiv F(t,0\,\vert \,x_0)$. 

In order to exploit previous results, we will first consider the above problem for $\mu=0$, i.e., for a collection of $N$ immortal walkers (quantities with hats will hereafter refer to immortal walkers). The probability that the target is annihilated by any of the walkers at time $t$ is characterized by the density $\hat{f}_N(t)$, which is in turn expressible in terms of quantities for the one-walker problem ~\cite{Lindenberg1983,Monasterio, Meerson15}:
\begin{align}
\hat{f}_N(t) &= N \hat{f}_1(t) \big[1-\hat{F}_1(t)\big]^{N-1}\nonumber \\
& = N \,\frac{x_0}{\sqrt{4\pi Dt^3}}\,\, e^{-x_0^2/4Dt}   \left[\mathrm{erf}\big(x_0/\sqrt{4Dt}\big)\right]^{N-1}\nonumber \\
&\simeq  N \frac{x_0}{\sqrt{4\pi Dt^3}}\,
\left(\frac{x_0}{\sqrt{\pi Dt}}\right)^{N-1},  \qquad t\to\infty,
\end{align}
where the third line follows from the $z\ll 1$ asymptotics of $\mathrm{erf}(z)$. A consequence of the fact that $\hat{f}_N(t)$ exhibits the algebraic long-time tail $\hat{f}_N(t)\sim t^{-(N+2)/2}$ is that the average search time 
\be
 \langle \hat{T}_N\rangle \!=\! \int_0^\infty \!\! dt\,t \,\hat{f}_N(t)   
\ee
(time taken by the fastest walker to reach the target) is divergent for $N\leq 2$, but finite for $N\geq 3$~\cite{Lindenberg80}.

In Ref.~\cite{Lindenberg1983}, the statistics of first-arrival times of $N\gg 1$ immortal walkers to the target was studied in detail. In particular, the first two moments of $\hat{T}_N$ were computed, and an asymptotic expansion with a (not fully correct) first subdominant term was provided. This problem was revisited in Ref.~\cite{Yuste1996}; the first subdominant term was amended and the next one (needed to estimate the variance) was also evaluated. To first order in the corrective terms, the final expression for $\langle \hat{T}_N \rangle$ reads~\cite{Yuste1996,Yuste2001} 
\begin{align}
\label{TNasy}
\langle \hat{T}_N\rangle= \frac{x_0^2}{4D\ln{\left(\lambda_0\, N\right)}}
\left\{
1+\frac{\frac{1}{2}\,\ln\left(\ln\left(\lambda_0 N\right)\right) -\gamma}{\ln \left(\lambda_0 N\right)}
\right\},
\end{align}
where $\lambda_0=1/\sqrt{\pi}$ and 
$\gamma\approx0.5772$ is the Euler-Mascheroni constant. Note that the corrective term remains non-negligible even for very large values of $N$. The approximate expression given by the leading term in Eq.~\eqref{TNasy} only, 
\be
\label{time0}
\langle \hat{T}_N \rangle \simeq \frac{x_0^2}{4 D\,\ln N},
\ee
agrees with the one obtained in Ref.~\cite{Meerson15} by a different method.

Turning now to the case of mortal walkers, for $N$ identical searchers all starting from $x_0$, the probability $p$ that at least one of them eventually hits the target is given by the interplay between the intrinsic mean lifetime $1/\mu$ and the diffusive time scale $\tau_D=x_0^2/4D$. Meerson and Redner~\cite{Meerson15} conclude that if $N$ is large enough, $p$ becomes non-negligible even for a relatively high mortality. Reaching the target is no longer guaranteed, but when $M\equiv \sqrt{\mu \tau_D}\gg 1$, one has for the conditional MFPT~\cite{Meerson15}
\be
\label{tNa}
\langle T_N\rangle = \frac{\int_0^\infty t \,f_N(t)\, dt}{\int_0^\infty f_N(t)\, dt}\simeq
\frac{\tau_D}{M}=\frac{x_0}{\sqrt{4D\mu}}.
\ee
The result~\eqref{tNa} is consistent with Eq.~\eqref{continuummfpt} for the MFPT of a \emph{single} mortal searcher.  Thus, in the high-mortality regime the MFPT is predominantly influenced by the fastest searcher only, which renders the other searchers superfluous unless their number is extremely large. In this latter case, when $N\!\gg\!\sqrt{M} e^{2M}$ (but with $M$ also large),  one recovers  Eq.~\eqref{time0}~\cite{Meerson15}. Therefore, in the large-$N$ limit, even a relatively high mortality becomes ultimately irrelevant for the computation of the MFPT. Thus, $\langle T_N\rangle \approx \langle \hat{T}_N\rangle$ in this limit. Consequently, although the search time (\ref{time0}) decays logarithmically with $N$, in the problem of oocyte fertilization it still entails a reduction by a factor of around $20$ with respect to the characteristic diffusion time $\tau_D$ because of the typically large number of human sperm ($N=3 \times 10^8$)~\cite{Meerson15}.

Finally, we note that an extension of the above $N$-walker problem to the case of a position-dependent death rate $\mu(x)$ was recently studied by Toste and Holcman~\cite{Toste2023}.

\subsection{Longest first-passage time}

The problem of estimating the moments of the time $\hat T_{j,N}$ taken by the $j$-th fastest searcher among a collection of $N\gg 1$ \emph{immortal} searchers to reach a predefined set of targets was addressed in Refs.~\cite{Lindenberg1983,Yuste1996,Yuste2001} for several initial distributions of searchers and different target arrangements in $d=1$. Variants of the same problem with different embedding geometries and target distributions were later considered, e.g., $d$-dimensional Euclidean lattices with absorbing hyper-spherical surface~\cite{Yuste2000,Yuste2001b}, $d$-dimensional Euclidean media filled with random traps~\cite{Yuste2001c}, disordered media filled with random traps~\cite{Yuste2003}, disordered media with an absorbing hyperspherical (Euclidean or chemical) surface in~\cite{Acedo2002}, and fractal lattices~\cite{Yuste1997,Yuste1998}. In particular, for $N$ immortal walkers starting all at $x_0$ and a point target at $x=0$, one has~\cite{Yuste1996}
\begin{equation}
\langle \hat T_{j,N}\rangle=\langle \hat T_{N}\rangle+ \frac{x_0^2}{4D\ln^2\left(\lambda_0 N\right)}\, H_{j-1}.
\end{equation}
$H_k=\sum_{r=1}^k r^{-1}$  is the $k$-th harmonic number, and the expression for  $\langle \hat T_{N}\rangle \equiv \langle \hat T_{1,N} \rangle $ is given by Eq.~\eqref{TNasy}.

Beyond a number of biological systems where the shortest FPT plays a central role, Lawley and Johnson~\cite{Lawley23} point out that in other instances the quantity of interest is the search time of the \emph{slowest} walker(s), i.e., the longest FPT, ``as it can define the termination of a process or perhaps the exhaustion of a supply''. In this context, Lawley and Johnson deal extensively with the onset of menopause, which occurs when the number of primordial follicles in the ovarian reserve drops from an initial value of around $5\times 10^5$ to approximately $10^3$ (see~\cite{Lawley23} and references therein).

Using extreme value statistics, Lawley and Johnson recently proved ~\cite{Lawley23} that when the survival probability $S(t)$ of an individual \textit{immortal} searcher exhibits the behavior $S(t)\sim (\zeta t)^{-\eta}$ for $t\to\infty$, then for $N\to\infty$ \textit{mortal} searchers and a fixed $k$, the MFPT of the $N-k$-th searcher behaves as follows:
\be
\label{TNkN}
\langle T_{N-k,N} \rangle
=\mu^{-1}\big(\ln N-(\eta+1)\ln\left(\ln N\right)+\ln \mathcal{A}+\gamma-H_{k}+o(1)\big), 
\ee
where $\mathcal{A}=\eta (\mu/\zeta)^\eta/\int_0^\infty [1-S(t')]\mu e^{-\mu t'}\,dt'$. In particular, for our canonical example of $N$ mortal walkers starting at $x_0$ with a target at $x=0$, one has $S(t)= \mathrm{erf}[x_0/(4Dt)]$, $\zeta=\pi D/x_0^2$, $\eta=1/2$ and $\mathcal{A}=\sqrt{\mu x_0^{2}/(4\pi D)}  \exp\left({\sqrt{\mu x_0^{2}/D}}\right)$.

\section{The target problem with anomalous diffusion}
\label{searchproblems}

Problems with anomalous diffusion and mortality have been addressed by many authors. Theoretical approaches rely often on (separable) CTRW models and the corresponding fractional diffusion equations, as they lend themselves well to the inclusion of chemical reactions in general and of decay processes in particular~\cite{Yuste06, Sokolov06, Henry06, Zoia08, Campos09, Abad10, Abad12, Abad15, Angstmann13, Stage17, Lawley20, Bresloff23, Bresloff23b}. We note, however, the case of non-separable walks~\cite{Campos15, Fedotov15} has also been addressed, as well as that of a decay process coupled to the statistics of jumps~\cite{Hornung05, Shkilev11, Abad13b} or to the territory explored~\cite{Benichou14, Benichou16}.

In the remainder of this section, we focus on the survival probability of an immobile target surrounded by one or more subdiffusively moving evanescent searchers. These searchers follow the statistics of an uncoupled CTRW with a jump length pdf of finite variance and a waiting-time pdf displaying the long-time behavior $t^{-1-\gamma}$ (with $0<\gamma<1$)~\cite{Abad12, Abad13, Abad15}. On sufficiently long spatial and temporal time scales, this yields a fractional diffusion equation with anomalous diffusion coefficient $D_\gamma$
($D_1\equiv D$ recovers the case of normal diffusion)~\cite{Metzler00}.

\subsection{The one-searcher problem}
\label{singletrapproblem}

Our setting consists of an impenetrable, hyperspherical target with radius $R$ placed at the origin $\mathbf{x}=0$ of a $d$-dimensional volume $V$, and an evanescent subdiffusive point searcher located inside $V$ at an initial position $\mathbf{x}_0$ ($x_0>R$). The target is spherical in three dimensions, circular in two dimensions, and a line in one dimension. Unless stated otherwise, the searcher's survival probability is assumed to decrease exponentially when the target is absent, i.e.,  $p_\mu(t)=\mu e^{-\mu t}$ . Since the searcher may evanesce as it moves, its pdf $p(\mathbf{x},t\vert\,\mathbf{x}_0)$ follows the subdiffusion-decay equation~\cite{Yuste10,Abad12} 
\be
\frac{\partial p(\mathbf{x},t \vert\,\mathbf{x}_0)}{\partial t}
=e^{-\mu t} D_{\gamma}
~_{0}{\cal D}_t^{1-\gamma} e^{\mu t} \nabla_{\mathbf{x}}^2 p(\mathbf{x},t \vert\,\mathbf{x}_0)
-\mu p(\mathbf{x},t \vert\,\mathbf{x}_0)
\label{subdiffusion-decay}
\ee
 where $~_{0}{\cal D}_t$ denotes the Riemann-Liouville fractional derivative\footnote{Strictly speaking, the derivation of  Eq.~\eqref{subdiffusion-decay} involves the so-called Gr\" unwald-Letnikov derivative, which is equivalent to the Riemann-Liouville one in the present case of a sufficiently smooth function at $t=0$  (see, e.g., Eqs. (2.255), (2.248) and (2.240) in Ref.~\cite{Podlubny99}).}
 \be
 \,_0{\cal D}_t^{1-\gamma}p(\mathbf{x},t \vert\,\mathbf{x}_0)\equiv \frac{1}{\Gamma(\gamma)} \frac{\partial}{\partial t}  \int_0^t dt'\, \frac{p(\mathbf{x},t \vert\,\mathbf{x}_0)}{(t-t')^{1-\gamma}}.
 \ee
We assume that the searcher destroys the target instantaneously upon contact and seek to compute the survival probability $Q_{1}(\mathbf{x}_0,t;R)$ of the target. Since the subsequent fate of the searcher is immaterial for this purpose, we will assume for convenience that it also dies instantly upon hitting the target (the particular case $d=1$, $R=0$, and $\gamma=1$ recovers the OSTP studied in Sect.~\ref{FPTsingle}). The searcher's survival probability $Q_{1,s}(\mathbf{x}_0,t;R)=\int p(\mathbf{x},t\vert\,\mathbf{x}_0) d\mathbf{x}$ is related to $Q_1$ via the equation~\cite{Abad12,Abad13}:
\be
Q_1(\mathbf{x}_0,t;R) = Q_{1,s}(\mathbf{x}_0,t;R)+\mu \int_0^t Q_{1,s}(\mathbf{x}_0,t^\prime; R) dt^\prime.
\label{gettingthere}
\ee
Thus, the target has a higher survival probability than the searcher at any time, since the latter may evanesce spontaneously before being able to hit the target [cf. Eq.~\eqref{subdiffusion-decay}]. To compute $Q_{1,s}(\mathbf{x}_0,t;R)$, one must solve the boundary value problem
\begin{subequations}
\begin{align}
\frac{\partial Q_{1,s}(\mathbf{x}_0,t;R)}
{\partial t}&=e^{-\mu t} D_{\gamma}
~_{0}{\cal D}_t^{1-\gamma} e^{\mu t} \nabla_{\mathbf{x}_0}^2 
Q_{1,s}(\mathbf{x}_0,t;R) \nonumber \\
&-\mu Q_{1,s}(\mathbf{x}_0,t;R),
\label{geneqQ*1arbrho2} \\
Q_{1,s}(\mathbf{x}_0,0;R) &=1, \\
Q_{1,s}(R,t;R) &=0, \\
Q_{1,s}({x}_0\to\infty,t;R)&  =e^{-\mu t}.
\end{align}
\end{subequations}
For $d=1,3$ one obtains the following long-time expression~\cite{Abad13}:
\be
{Q}_1(x_0,\infty;R)-{Q}_1(x_0,t;R)\sim -
\left(\frac{R}{x_0}\right)^{(d-1)/2}\frac{1}{\Gamma(1-\frac{\gamma}{2})}\,\frac{x_0-R}{\mu^{1+\frac{\gamma}{2}}
\,\ell_\gamma}\,t^{-1-\frac{\gamma}{2}}e^{-\mu t},
\ee
where $\ell_\gamma=\left(4D_\gamma /\mu^\gamma\right)^{1/2}$ can be regarded as the typical distance explored by the searcher before dying spontaneously and
\be
\label{stsp}
{Q}_1(x_0,\infty;R)=1-\left(\frac{R}{x_0}\right)^{(d-1)/2} \, e^{-2(x_0-R)/\ell_\gamma}.
\ee
In $d=2$, the long-time decay takes a more complicated form:
\begin{align}
\label{asanint}
Q_1(x_0,t;R) \sim &~Q_1(x_0, \infty;R)+\frac{2}{\gamma}\ln \left(\frac{x_0}{R}\right) \frac{e^{-\mu t}}
{\mu t\ln^2(\alpha_\gamma t)}
\end{align}
with $\alpha_\gamma=(4D_\gamma/R^2)^{1/\gamma}$ and
\be
Q_1(x_0, \infty;R)=1-\frac{K_0(2x_0/\ell_\gamma)}
{K_0(2R/\ell_\gamma)}.
\ee
Thus, in $d=1,2$ the target has a non-zero probability of eternal survival, as opposed to the standard case of an immortal searcher ($\mu=0$). In this latter case, the target death would be certain because of the recurrence of the searcher's walk.

\subsection{The many-searcher problem}

The results for the one-searcher target problem provide the starting point to address the many-searcher problem. Specifically, consider a hyperspherical target of radius $R$ centered at $\mathbf{x}=0$ and surrounded by an ensemble of $N_0$ non-interacting evanescent point searchers. These mortal searchers are randomly distributed in a hyperspherical volume $V$ around $\mathbf{x}=0$, and each of them satisfies the diffusion-decay equation~\eqref{subdiffusion-decay}. As in the one-searcher problem, we assume that if any searcher hits the target, both are instantaneously annihilated. We again focus on the survival probability $Q(t;R)$ of the target, i.e., the probability that it has not been hit by any searcher up to time $t$.

Because of the statistical independence of the searchers' trajectories, the solution can be expressed in terms of its counterpart for the one-searcher problem:
\begin{eqnarray}
\label{Q*Ta}
  Q(t;R)= \left[\frac{1}{V} \int_{ {x}_0>R}
  Q_1(\mathbf{x}_0,t;R)\, d\mathbf{x}_0\right]^{N_0}.
\end{eqnarray}
For simplicity, we take the thermodynamic limit $N_0\to\infty,V\to \infty$; in other words, we fix the initial global density of searchers, i.e.,  $\rho_0\equiv \rho(0)=\lim_{N_0,V \to\infty} N_0/V$. In this limit, from the solution for $Q_1$ in an infinite volume, one finds 
\be
\label{Qsigma}
Q(t;R)=\exp\left\{-\rho_0 R^d \sigma(t,R) \right\}
\ee
with
\be
 \sigma(t;R)\equiv
 \frac{1}{R^d} \int_{ {x}_0>R}[1-Q_1(\mathbf{x}_0,t;R)]\,
 d\mathbf{x}_0.
 \label{sigmatR}
\ee
As pointed out in Refs.~\cite{Abad13b} and~\cite{Abad15}, the quantity $R^d \sigma(t,R)$ is directly related to the average volume swept by a \emph{single} mortal walker up to time $t$. Long-time asymptotic expressions for $\sigma(t,R)$ were obtained in Refs.~\cite{Abad12} and ~\cite{Abad13}. The general expression for the final value in $d$ dimensions is 
\be
\label{spalld}
\sigma(\infty;R) =S_d \frac{\ell_\gamma}{2R}\frac{K_{d/2}(2R/\ell_\gamma)}{K_{d/2-1}(2R/\ell_\gamma)},
\ee
where $S_d=2\pi^{d/2}/\Gamma(d/2)$ stands for the surface of a $d$-dimensional hypersphere of unit radius and $K_\nu$ is the $\nu$-th order modified Bessel function of the second kind. 
Thus, even though we have an infinite initial number of searchers, the target has a non-zero chance of eternal survival in all dimensions. Thus, in a finite fraction of realizations, a sufficiently fast evanescence process may kill all the searchers before any of them is able to hit the target. In the absence of such a process, the target is of course eventually killed with certainty in all spatial dimensions~\cite{Yuste06}.

To close, we mention that the problem with a power-law decaying density of searchers $\rho(t)\propto t^{-\beta}$ is also interesting; depending on the values of $\beta$ and $\gamma$, the final survival probability of the target $Q(\infty;R)$ is finite or not. We will not deal with this problem here for reasons of space, but the details can be found in Ref.~\cite{Abad13}.

\section{The target problem in an evolving domain}

A rubber band that stretches/shrinks or the surface of a balloon being inflated/deflated are elementary examples of evolving media. Some motivation for studying diffusion in expanding media comes e.g. from the fields of developmental biology (diffusion-mediated formation of growing biological structures) and cosmology (diffusion of cosmic rays in the expanding universe), see e.g.~\cite{Yuste16,BookChapter19,Abad2020} and references therein. In particular, the search problems addressed in Sect.~\ref{searchproblems} can now be considered in an expanding (or contracting) domain.
 
In the case of a shrinking domain, the searcher concentration grows as the volume decreases, and this may compensate for the particle loss due to evanescence. Therefore, a competition between both mechanisms settles. For simplicity, we illustrate this competition via the normal diffusive case ($\gamma=1$) on the real line ($d=1$).

\subsection{The one-searcher problem}

As before, we consider the target problem with a single evanescent point searcher, but now each length element $dy$ of the real line grows/shrinks (homogeneously) in the course of time as $dy=a(t)\,dx$, where $dx$ denotes its initial length and $a(t)$ is the so-called scale factor, which takes the initial value $a(0)=1$~\cite{Yuste16}. Thus, a point on the line with initial coordinate $x$ (comoving coordinate) is located at the physical coordinate $y=a(t)\, x$ after a time $t$.

We now place a target (i.e., a segment with absorbing endpoints) centered at $y=x=0$. We assume that the target size follows the evolution law of the one-dimensional domain, i.e., $R(t)=a(t) R$, where $R$ denotes the initial ``radius'' (half-length of the target). The searcher starts at $y_0=x_0$ with $x_0>R$ and performs random jumps drawn from a stationary pdf $\lambda(y)$ with finite variance. In $x$-space, the corresponding jump length pdf becomes time-dependent $\lambda'(x,t)$, and so does the diffusivity. Between consecutive jumps, the searcher sticks rigidly to the expanding/shrinking length element on which it rests; thus, in $y$-space, the searcher experiences a deterministic drift while still waiting to jump. On the other hand, between $t$ and $t+dt$, the searcher has a non-zero probability $\mu dt$ to spontaneously disappear. The corresponding reaction-diffusion equation in $x$-space then is~\cite{Abad2020}:
\be
\frac{\partial p(x,t\vert\, x_0)}{\partial t}
=\frac{D}{a^2(t)} \frac{\partial^2 p(x,t\vert\, x_0)}{\partial x^2}
-\mu p(x,t\vert\, x_0).
\label{expanding-diffusion-decay}
\ee
From here, one can easily compute the searcher's survival probability $Q_{1,s}(x_0,t)=\int p(x,t \vert\, x_0) dx$. The solution of the corresponding boundary value problem is 
\be
Q_{1,s}(x_0,t)=e^{-\mu t} \hat{Q}_{1,s}(x_0,t) =e^{-\mu t} \mathrm{erf}\left(\frac{x_0-R}{\sqrt{4D \tau(t)}}\right)
\ee
with the so-called Brownian conformal time $\tau(t)=\int_0^t dt'/a^2(t')$~\cite{Yuste16, Abad2020}. The conformal time $\tau$ accounts for the effect of the domain deformation; it grows superlinearly (sublinearly) in time in a shrinking (expanding) domain. 

In the present case of an evolving domain, the survival probability of the target $Q_1(x_0,t)>Q_{1,s}(x_0,t)$ is still given by Eq.~\eqref{gettingthere}. Depending on the form $a(t)$, the integral on the right-hand side (which quantifies the difference between both survival probabilities) may or may not be explicitly evaluated. In some cases, the asymptotic value $Q_1(x_0,\infty)$ and the long-time approach to it can be computed via Laplace transforms and Tauberian theorems. In any case, for fixed $x_0$ and $t$, $Q_1(x_0,t)$ is larger (smaller) on the expanding (shrinking) line than its counterpart for a static domain. 

\subsection{The many-searcher problem}

Eqs.~\eqref{Q*Ta},~\eqref{Qsigma}, and~\eqref{sigmatR} still hold in $x$-space. From Eqs.~\eqref{sigmatR} and~\eqref{gettingthere}, one finds
\be
\frac{\partial \sigma(t;R)}{\partial t}=-\frac{1}{R}\int_{x_0>R}\left[\frac{\partial Q_{1,s}}{\partial t}+\mu e^{-\mu t} Q_{1,s}\right] dx_0.
\ee
We now use Eq.~\eqref{expanding-diffusion-decay} and change the ``volume integral'' into a ``surface integral''
\be
\frac{\partial \sigma(t;R)}{\partial t}= \frac{2D}{a^2(t)} \frac{e^{-\mu t}}{R} \left.\frac{\partial \hat{Q}_{1,s}(x_0, t;R)}{\partial x_0}\right|_{x_0=R}
\ee
via Gauss's theorem~\cite{Abad12}. Integration from 0 to $t$ yields
\be
\sigma(t ; R)= \frac{2 D}{R} \int_{0}^{t}\frac{1}{a^2(t')}\left.\frac{\partial \hat{Q}_{1,s}\left(x_0, t^{\prime} ; R\right)}{\partial x_0}\right|_{x_0=R} e^{-\mu t'}\,dt'
\ee
(recall that $\sigma(0;R)=0$ $\left[Q(0;R)=1\right]$). Finally, the explicit form of $\hat{Q}_{1,s}$ gives
\be
\sigma(t; R)= R^{-1}\sqrt{\frac{4D}{\pi}} \int_{0}^{t}\frac{1}{a^2(t')}
\frac{e^{-\mu t'}}{\sqrt{\tau(t')}}\, dt'.
\ee
In particular, for an exponentially shrinking domain $a(t)=e^{Ht}$ with 
$H<0$ [$\tau(t)=H^{-1} e^{-Ht}\mathrm{sinh}(Ht)$], one sees that $Q(\infty,R)$ remains finite as long as $\mu > |H|$, but tends to zero otherwise. For an expanding line ($H>0$), the target has a finite chance of eternal survival even when the searchers are immortal (see Sect. 3.3.2 in~\cite{BookChapter19}).

\section{Escape}

Escape from a domain with reflecting boundaries through a narrow target is an important topic in biology~\cite{Holcman14, Holcman2015}. In this context, inactivation or degradation of the entity diffusing towards the target~\cite{Holcman2005} often competes on time scales comparable or shorter than those of pure diffusion, and must therefore be accounted for to compute rates of binding, etc. In Ref.~\cite{Grebenkov17}, Grebenkov showed that the mean escape time (here identified as the MFPT) and the target absorption probabilities can be related to their counterparts for immortal walkers. Below we sketch his derivation.  

Let $\hat{Q}(\mathbf{x}_0,t)$ denote the survival probability, i.e., the probability that an \emph{immortal} walker started at a point $\mathbf{x}_0\in\Omega$ has not left a confining domain $\Omega \subset \R^d$ through an escape region $\Gamma$ on the boundary $\pa$. The corresponding FPT density is
\be
\label{eq:rho}
\hat{\rho}(\mathbf{x}_0,t) = - \frac{\partial\hat{Q}(\mathbf{x}_0,t) }{\partial t} . 
\ee
Note that, if $\tilde{\hat{Q}}(\mathbf{x}_0,s)$ is the Laplace-transformed survival probability,
then the MFPT is simply $\tilde{\hat{Q}}(\mathbf{x}_0, 0)$. For \emph{mortal} walkers, assuming as before that the walker's mortality and the diffusion process are independent, one has that its survival probability (i.e., the probability that the walker has neither evanesced nor left the domain $\Omega$ up to time $t$) is simply $Q(\mathbf{x}_0,t)= e^{-\mu t} \hat{Q}(\mathbf{x}_0,t)$. Thus, in Laplace space, $\tilde{Q}(\mathbf{x}_0,s) = \tilde{\hat{Q}}(\mathbf{x}_0, s+\mu)$, and the MFPT for mortal walkers then takes the form
\be
\label{eq:MFET}
\langle T(\mathbf{x}_0) \rangle = \tilde{Q}(\mathbf{x}_0,0) = \tilde{\hat{Q}}(\mathbf{x}_0,\mu).
\ee 
There is a connection between the MFPT and the probability $H(\mathbf{x}_0)$ that a mortal walker started from $\mathbf{x}_0$ leaves the confining domain before dying, namely,   
\be
\label{eq:Hexit}
H(\mathbf{x}_0) = \int\limits_0^\infty dt\, e^{-\mu t} \, \hat{\rho}(\mathbf{x}_0, t) = 1 - \mu \tilde{\hat{Q}}(\mathbf{x}_0,\mu) = 1 - \mu \langle T(\mathbf{x}_0) \rangle.   
\ee
For low mortality ($\mu \langle \hat{T}(\mathbf{x}_0) \rangle \ll 1$), the Taylor expansion of $\tilde{\hat{Q}}(\mathbf{x}_0,s)$ yields
\be
\label{eq:MFPT_small}
\langle T(\mathbf{x}_0) \rangle = \langle \hat{T}(\mathbf{x}_0) \rangle - \frac12 \mu 
 \langle \hat{T^2}(\mathbf{x}_0) \rangle+ O(\mu^2).
\ee
From Eq. (\ref{eq:Hexit}), we then find
\be
\label{eq:Hmu_small}
H(\mathbf{x}_0) = 1 - \mu \langle \hat{T}(\mathbf{x}_0) \rangle + \frac12 \mu^2 
\langle \hat{T^2}(\mathbf{x}_0) \rangle+ O(\mu^3) .
\ee
We are thus led to the analysis of moments of the FPT for immortal walkers:  
\be
\label{eq:tau_moments}
\langle \hat{T^n}(\mathbf{x}_0) \rangle = (-1)^n \left.\frac{\partial^n}{\partial \mu^n} H( \mathbf{x}_0) \right|_{\mu = 0} .
\ee
Exact asymptotic formulas for the MFPT of immortal walkers are known for some 
simple domains~\cite{Holcman14}. E.g., for $d=3$, $\langle \hat{T}(\mathbf{x}_0) \rangle \simeq  |\Omega|/(4DR \epsilon) + O(\ln \epsilon)$, where $R = \sqrt{|\pa|}$ is the characteristic size (``radius'') of the domain, and $\epsilon = (|\Gamma|/|\pa|)^{1/2} \ll 1$ is the square root of the area of the escape region normalized by the area of the boundary.
In Ref.~\cite{Grebenkov17}, a thorough study of the exit probability and the MFPT for small, moderate, and large death rates can be found. 

A variant of the above problem concerns the statistics of the escape time for $N$ walkers initially uniformly distributed throughout the confining domain~\cite{Grebenkov20}. Another situation of interest refers to diffusion processes switching between $d=3$ and $d=2$ (Adam-Delbr\"uck scenario) [see Ref.~\cite{Grebenkov22} for a recent treatment with one and $N$ walkers]. The effect of mortality in these two settings is still an open question.

\section{Summary and conclusion}

Mortality changes drastically the properties of Brownian motion inasmuch as it induces a stronger localization and promotes quasiballistic behaviour. Only sufficiently short trajectories have a chance to reach a target, as illustrated by the canonical example of one or many particles diffusing on the real half-line with a target placed at the origin: the conditional MFPT is finite and linear in the initial target-searcher separation; mortal random walks become non-recurrent already in one dimension, and quantities like the mean number of distinct sites visited converge to a finite value.

Resetting of an individual mortal walker eliminates long trajectories which lead to premature death before finding the target. On the other hand, an excessively large resetting decreases the search efficiency. The trade-off between these two effects leads to an optimal resetting rate induced by the mortality constraint. 

The order statistics of a collection of $N$ mortal walkers searching for the target depends strongly on how the mortality time scale compares to the diffusive one. For high mortality, the MFPT of the fastest searcher is identical with the one obtained in the one-searcher problem; for low mortality, the leading term of this quantity is the same as for immortal walkers and thus proportional to $ 1/\ln{N}$. In contrast, the MFPT of the slowest searchers turns out to grow as $\ln{N}$ in the large-$N$ limit. 

The target problem for the half-line can also be solved in the case of subdiffusive searchers. Both for a single searcher and for many (randomly distributed) searchers, the target has (as in the case of normal diffusion) a finite probability of eternal survival. The asymptotic decay of the target's survival probability to this finite value is given by the intrinsic survival probability of the mortal searcher $e^{-\mu t}$ multiplied by slower decaying terms (power-law, logarithmic) that depend on the searcher’s anomalous diffusion exponent.

Another variant of the one-dimensional target problem is obtained by introducing a homogeneous expansion (contraction) of the real line. For an exponential contraction, if the searchers' mortality rate is larger than the contraction rate, the target may survive forever even if the number of (randomly distributed) searchers is infinite. 

Finally, we have also reproduced some results for the conditional MFPT and the exit probability in the narrow escape problem for mortal walkers. Explicit expressions follow from the formalism for immortal walkers in the low-mortality limit. 
 
Many of the results presented here are recent, which shows that research on mortal walkers is very much alive. In view of the number of possible extensions and the wealth of applications, we expect that a good deal of new findings relevant for both theory and experiments will be reported in the near future.

\section{Acknowledgements}

We acknowledge financial support from Grant PID2020-112936GB-I00 funded by MCIN/AEI/10.13039/501100011033, and from Grant IB20079 funded by Junta de Extremadura (Spain) and by ERDF ``A way of making Europe''.



\begin{thebibliography}{99.}

 \bibitem{Nicolaysen57} Nicolaysen, L.O.: Solid diffusion in radioactive minerals and the measurement of absolute age. Geochim. Cosmochim. Acta \textbf{11}, 41--59 (1957)

 \bibitem{Silverman17} Silverman, M.P.: Brownian Motion of Decaying Particles: Transition Probability, Computer Simulation, and First-Passage Times. J. Mod. Phys. \textbf{8}, 1809--1849 (2017)

\bibitem{Meerson15} Meerson, B., Redner, S.: Mortality, Redundancy, and Diversity in Stochastic Search. Phys. Rev. Lett. \textbf{114}, 198101 (2015) 

\bibitem{Abad15b} Abad, E., Kozak, J.J.: Competing reaction processes on a lattice as a paradigm for catalyst deactivation. Phys. Rev. E \textbf{91}, 022106 (2015)

\bibitem{Sano79} Sano, H., Tachiya, M.: Partially diffusion controlled recombination. J. Chem. Phys. \textbf{71}, 1276--1282 (1979)

\bibitem{Seki05} Seki, K., Murayama, M., Tachiya, M.: Dispersive photoluminescence decay by geminate recombination in amorphous semiconductors. Phys. Rev. B \textbf{71}, 235212 (2005) 

\bibitem{Yuste2013} Yuste, S.B., Abad, E., Lindenberg, K.: Exploration and Trapping of Mortal Random Walkers. Phys. Rev. Lett. \textbf{110}, 220603 (2013)

\bibitem{Grebenkov17} Grebenkov, D.S., Rupprecht, J.-F.: The escape problem for mortal walkers. J. Chem. Phys. \textbf{146}, 084106 (2017)

\bibitem{Bonner87} Bonner, R.F., Nossal, R., Havlin, S., Weiss, G.H.: Model for photon migration in turbid biological media.  J. Opt. Soc. Am. A \textbf{4}, 423--432 (1987)

\bibitem{Wartlick09} Wartlick O., Kicheva A., González-Gait\'an, M., Morphogen gradient formation. Cold Spring Harb. Perspect. Biol. \textbf{1}, a001255 (2009)

\bibitem{Yuste10} Yuste, S.B., Abad, E., Lindenberg, K.: A reaction-subdiffusion model of morphogen gradient formation.  Phys. Rev. E \textbf{82}, 061123 (2010) 

\bibitem{Kolomeisky00} Kolomeisky, A.B., Fisher, M.E.: Periodic sequential kinetic models with jumping, branching and deaths. Physica A \textbf{279}, 1--20 (2000)

\bibitem{Holcman2015} Holcman, D., Schuss, Z.: Stochastic Narrow Escape in Molecular and Cellular Biology. Springer, New York (2015)

\bibitem{Ma20} Ma, J., Do, M., Le Gros, M.A., Peskin, C.S., Larabell, C.A., Mori, Y., Isaacson, S.A.: Strong intracellular signal inactivation produces sharper and more robust signaling from cell membrane to nucleus. PLoS Comput. Biol. \textbf{16}, e1008356 (2020)

\bibitem{Briantceva23} Briantceva, N., Parsani, M.: Impact of Evanescence Process on Three-Dimensional Sub-Diffusion based Molecular Communication Channel. IEEE Trans. Nanobioscience (2023) doi:10.1109/TNB.2023.326010

\bibitem{Lawley21} Lawley, S.D.: The Effects of Fast Inactivation on Conditional First Passage Times of Mortal Diffusive Searchers. SIAM J. Appl. Math. \textbf{81}, 1--24 (2021)

\bibitem{Bala19} Balakrishnan, V., Abad, E., Abil, T., Kozak, J.J.: First-passage properties of mortal random walks: Ballistic behavior, effective reduction of dimensionality, and scaling functions for hierarchical graphs. Phys. Rev. E \textbf{99}, 062110 (2019)

\bibitem{Grebenkov22} Grebenkov, D.S., Metzler, R., Oshanin, G.: 
Search efficiency in the Adam-Delbr\" uck reduction-of-dimensionality scenario versus direct diffusive search. New J. Phys. \textbf{24}, 083035 (2022) 

\bibitem{Fedotov13} Fedotov, S., Falconer S.: Random death process for the regularization of subdiffusive fractional equations. Phys. Rev. E \textbf{87}, 052139 (2013)

\bibitem{Sikder21} Sikder, O., Averaging dynamics, mortal random walkers and information aggregation on graphs. J. Phys. Complex. \textbf{2}, 045005 (2021) 

\bibitem{Hughes95} Hughes, B.D., Random Walks and Random Environments. Volume 1: Random Walks, pp. 123-124. Clarendon Press, Oxford (1995) 

\bibitem{Yuste14} Yuste, S.B., Abad, E., Lindenberg, K.: Arrival Statistics and Exploration Properties of Mortal Walkers. In: Metzler, R., Oshanin, G., Redner, S. (eds.) First-Passage Phenomena and their Applications, pp. 1-20. World Scientific, Singapore (2014)

\bibitem{Lohmar2009} Lohmar, I., Krug, J.: Diffusion-Limited Reactions and Mortal Random Walkers in Confined Geometries. J. Stat. Phys. \textbf{134}, 307--336 (2009)  

\bibitem{Abad13b} Abad, E., Yuste, S.B., Lindenberg, K.: Evanescent continuous time random walks. Phys. Rev. E \textbf{88}, 062110 (2013)

\bibitem{Re15} Ré, M.A., Bustos, N.C.: Reaction rate in an evanescent random walkers system. Pap. Phys. \textbf{7}, 070003 (2015)

\bibitem{Esguerra17} Esguerra, J.P., Reyes, J.: Exploration properties of biased evanescent random walkers on a one-dimensional lattice. AIP Conf. Proc. \textbf{1871}, 050002 (2017)

\bibitem{Abad18} Abad, E., Abil, T., Santos, A., Kozak, J.J.: Random walks on lattices. Influence of competing reaction centers on diffusion-controlled processes. Physica A \textbf{511}, 336--357 (2018)

\bibitem{Wu20} Wu, Z., Xu G.: Mortal random walks on a family of treelike regular fractals with a deep trap. Int. J. Mod. Phys. B \textbf{34}, 2050031 (2020)

\bibitem{Meerson19} Meerson, B.: Mortal Brownian motion: Three short stories. Int. J. Mod. Phys. B \textbf{33}, 1950172 (2019)

\bibitem{Abad13} Abad, E., Yuste, S.B., Lindenberg, K.: Elucidating the Role of Subdiffusion and Evanescence in the Target Problem: Some Recent Results. Math. Model. Nat. Phenom. \textbf{8}, 100--113 (2013)

\bibitem{Berezhkovskii1989}
Berezhkovskii, A.M., Makhnovskii, Y.A., Suris, R.A.: Wiener sausage volume moments. J. Stat. Phys. \textbf{57}, 333–346 (1989)

\bibitem{Evans2011} Evans, M.R., Majumdar, S.N.: Diffusion with Stochastic Resetting. Phys. Rev. Lett. \textbf{106}, 160601 (2011)

\bibitem{Evans20} Evans, M.R., Majumdar, S.N., Schehr, G.: Stochastic resetting and applications. 
J. Phys. A: Math. Theor. \textbf{53} 193001 (2020) 

\bibitem{Belan18} Belan, S.: Restart Could Optimize the Probability of Success in a Bernoulli Trial. Phys. Rev. Lett. \textbf{120}, 080601 (2018)

\bibitem{Radice23} Radice, M.: Effects of mortality on stochastic search processes with resetting. Phys. Rev. E \textbf{107}, 024136 (2023)

\bibitem{Harbison04} Harbison, C.T., Gordon, D.B., Lee, T.I., Rinaldi, N.J., Macisaac, K.D., Danford, T.W., Hannett, N.M., Tagne, J.B., Reynolds, D.B., Yoo, J.\textit{ et al.}: Transcriptional regulatory code of a eukaryotic genome. Nature \textbf{431}, 99--104 (2004)

\bibitem{Schuss19} Schuss, Z., Basnayake, K., Holcman, D.: Redundancy principle and the role of extreme statistics in molecular and cellular biology. Phys. Life Revs. \textbf{28}, 52--79 (2019)

\bibitem{Wang08} Wang, S.P., Pei, W.J.: First passage time of multiple Brownian particles on networks with applications. Physica  A \textbf{387}, 4699--4708 (2008)

\bibitem{Lindenberg1983} Weiss, G.H., Shuler, K.E., Lindenberg, K.: Order statistics for first passage times in diffusion processes. J. Stat. Phys. \textbf{31}, 255--278 (1983)

\bibitem{Monasterio} Mejía-Monasterio, C., Oshanin, G., Schehr, G.: First passages for a search by a swarm of independent random searchers. J. Stat. Mech. P06022 (2011) 

\bibitem{Lindenberg80}  Lindenberg, K., Seshadri, V., Shuler, K.E., Weiss G.H.: Lattice random walks for sets of random walkers. First passage times. J. Stat. Phys. \textbf{23}, 11--25 (1980) 

\bibitem{Yuste1996} Yuste, S.B., Lindenberg, K.: Order statistics for first passage times in one-dimensional diffusion processes. J. Stat. Phys. \textbf{85}, 501--512 (1996) 

\bibitem{Yuste2001}
S. B. Yuste, Acedo, L.: Multiparticle trapping problem in the half-line. Physica A \textbf{297}, 321--336 (2001)

\bibitem{Toste2023} Toste, S., Holcman, D.: Extreme Diffusion with Point-Sink Killing Fields: Application to Fast Calcium Signaling at Synapses. SIAM J. Appl. Math. \textbf{84}, S269--S296 (2024)

\bibitem{Yuste2000} Yuste, S.B., Acedo, L.: Diffusion of a set of random walkers in Euclidean media. First passage times. J. Phys. A \textbf{33}, 507--512 (2000)

\bibitem{Yuste2001b} Yuste, S.B., Acedo, L., Lindenberg, K.: Order statistics for $d$-dimensional diffusion processes. Phys. Rev. E \textbf{64}, 052102 (2001)

\bibitem{Yuste2001c} Yuste, S.B., Acedo, L.: Order statistics of the trapping problem.  Phys. Rev. E \textbf{64}, 061107 (2001)

\bibitem{Yuste2003} Yuste, S.B., Acedo, L.: Order statistics of Rosenstock’s trapping problem in disordered media. Phys. Rev. E \textbf{68}, 036134 (2003)

\bibitem{Acedo2002} Acedo, L., Yuste, S.B.: Survival probability and order statistics of diffusion on disordered media. Phys. Rev. E \textbf{66}, 011110 (2002)

\bibitem{Yuste1997} Yuste, S.B.: Escape times of $N$ random walkers from a fractal labyrinth. 
Phys. Rev. Lett. \textbf{79}, 3565--3568 (1997)

\bibitem{Yuste1998} Yuste, S.B.: Order statistics of diffusion on fractals. Phys. Rev. E \textbf{57}, 6327--6333 (1998)

\bibitem{Lawley23} Lawley, S.D., Johnson, J.: Slowest first-passage times, redundancy, and menopause timing. J. Math. Biol. \textbf{86}, 90 (2023) 

\bibitem{Yuste06} Yuste, S.B., Ruiz-Lorenzo, J.J., Lindenberg, K.: Target problem with evanescent subdiffusive traps. Phys. Rev. E \textbf{74}, 046119 (2006) 

\bibitem{Sokolov06} Sokolov, I.M., Schmidt, M.G.W., Sagués, F.: Reaction-subdiffusion equations. Phys. Rev. E \textbf{73}, 031102 (2006)

\bibitem{Henry06} Henry, B.I., Langlands, T.A.M., Wearne, S.L.: Anomalous diffusion with linear reaction dynamics. Phys. Rev. E \textbf{74}, 031116 (2006)

\bibitem{Zoia08} Zoia, A.: Continuous-time random-walk approach to normal and anomalous 
reaction-diffusion processes. Phys. Rev. E \textbf{77}, 041115 (2008)

\bibitem{Campos09} Campos, D., Méndez, V.: Nonuniversality and the role of tails in reaction-subdiffusion fronts. Phys. Rev. E \textbf{80}, 021133 (2009)

\bibitem{Abad10} Abad, E., Yuste, S.B., Lindenberg, K.: Reaction-subdiffusion and reaction-superdiffusion equations for evanescent particles performing continuous-time random walks. Phys. Rev. E \textbf{81}, 031115 (2010)

\bibitem{Abad12} Abad, E., Yuste, S.B., Lindenberg, K.: Survival probability of an immobile target in a sea of evanescent diffusive or subdiffusive traps: A fractional equation approach. Phys. Rev. E \textbf{86}, 061120 (2012)

\bibitem{Abad15} Abad, E., Yuste, S.B., Lindenberg, K.: Fractional Reaction-Transport Equations Arising From Evanescent Continuous Time Random Walks. In: Abi Zeid Daou, R., Moreau, X. (eds.) Fractional Calculus: Theory. Nova Science Publishers, New York (2015).

\bibitem{Angstmann13} Angstmann, C.N., Donnelly, I.C., Henry, B.I.: Continuous Time Random Walks with Reactions Forcing and Trapping. Math. Model. Nat. Phenom. \textbf{8}, 17--27 (2013) 

\bibitem{Stage17} Stage, H.: Aging in mortal superdiffusive Lévy walkers. Phys. Rev. E \textbf{96}, 062150 (2017)

\bibitem{Lawley20} Lawley, S.: Anomalous reaction-diffusion equations for linear reactions. Phys. Rev. E \textbf{102}, 032117 (2020) 

\bibitem{Bresloff23} Bresloff, P.C.: Encounter-based reaction-subdiffusion model I: surface adsorption and the local time propagator. J. Phys. A Math. Theor. \textbf{56}, 435004 (2023)

\bibitem{Bresloff23b} Bresloff, P.C.: Encounter-based reaction-subdiffusion model II: partially absorbing traps and the occupation time propagator. 
J. Phys. A Math. Theor. \textbf{56}, 435005 (2023)

\bibitem{Campos15} Campos, D., Abad, E., M\'endez, V., Yuste, S.B., Lindenberg, K.: Optimal search strategies for space-time coupled random walkers with finite lifetimes. Phys. Rev. E \textbf{91}, 052115 (2015)

\bibitem{Fedotov15} Fedotov, S., Tan, A., Zubarev, A.: 
Persistent random walk of cells involving anomalous effects and random death.
Phys. Rev. E \textbf{91}, 042124 (2015)

\bibitem{Hornung05} Hornung, G., Berkowitz, B., Barkai, N.: 
Morphogen gradient formation in a complex environment: An anomalous diffusion model. Phys. Rev. E \textbf{72}, 041916 (2005) 

\bibitem{Shkilev11} Shkilev, V.P.: Subdiffusion with the disappearance of particles at the time of a jump. J. Exp. Theor. Phys. \textbf{112}, 1071--1076 (2011)

\bibitem{Benichou14} Bénichou, O., Redner, S.: Depletion-Controlled Starvation of a Diffusing Forager. Phys. Rev. Lett. \textbf{113}, 238101 (2014)

\bibitem{Benichou16} Bénichou, O., Chupeau, M., Redner, S.: Role of depletion on the dynamics of a diffusing forager. J. Phys. A: Math. Theor. \textbf{49}, 394003 (2016)

\bibitem{Metzler00} Metzler, R., Klafter, J.:
The random walk's guide to anomalous diffusion: a fractional dynamics approach.
Phys. Rep. \textbf{339}, 1--77 (2000) 

\bibitem{Podlubny99} Podlubny, I.: Fractional Differential Equations: An Introduction to Fractional Derivatives, Fractional Differential Equations, to Methods of Their Solution and Some of Their Applications. Academic Press, San Diego (1999)

\bibitem{Yuste16} Yuste, S.B., Abad, E., Escudero, C.: Diffusion in an expanding medium: Fokker-Planck equation. Green’s function, and first-passage properties. Phys. Rev. E \textbf{94}, 032118 (2016)

\bibitem{BookChapter19} Abad, E., Escudero, C., Le Vot, F., Yuste, S.B.: First-Passage Processes and Encounter-Controlled Reactions in Growing Domains. In: Lindenberg, K., Metzler, R., Oshanin, G. Chemical Kinetics: Beyond the Textbook. World Scientific, New Jersey (2019)

\bibitem{Abad2020} Abad, E., Angstmann, C.N., Henry, B.I., McGann, A.V., Le Vot, F., Yuste, S.B.: Reaction-diffusion and reaction-subdiffusion equations on arbitrarily evolving domains. Phys. Rev. E \textbf{102}, 032111 (2020) 

\bibitem{Holcman14} Holcman, D., Schuss, Z.: The Narrow Escape Problem. SIAM Rev. \textbf{56}, 213--257 (2014)

\bibitem{Holcman2005} Holcman, D., Marchewka, A., Schuss, Z.: Survival probability of diffusion with trapping in cellular neurobiology. Phys. Rev. E \textbf{72}, 031910 (2005)

\bibitem{Grebenkov20} Grebenkov, D.S., Metzler, R., Oshanin, G.: From single-particle stochastic kinetics to macroscopic reaction rates: fastest first-passage time of $N$ random walkers, New J. Phys. \textbf{22}, 103004 (2020) 

\end{thebibliography}
\end{document}